\documentclass[10pt,letterpaper,twocolumn]{article}
\usepackage[english]{babel}
\usepackage{graphicx}

\newcommand{\alt}{\mathbin{\lower 3pt\hbox
   {$\rlap{\raise 5pt\hbox{$\char'074$}}\mathchar"7218$}}}
\newcommand{\agt}{\mathbin{\lower 3pt\hbox
   {$\rlap{\raise 5pt\hbox{$\char'076$}}\mathchar"7218$}}}

\begin{document}

\setcounter{footnote}{0}
\setcounter{equation}{0}
\setcounter{figure}{0}
\setcounter{table}{0}

\title{\large\bf Conductance distribution near the
Anderson transition}

\author{\small I. M. Suslov  \\
\small P.L.Kapitza Institute for Physical Problems,  \\
\small 119334 Moscow, Russia  \\
\small E-mail: suslov@kapitza.ras.ru\\
 {}\\
\parbox{155mm}{\footnotesize \,Using a modification of the
Shapiro approach, we introduce the two-parameter family of
conductance distributions $W(g)$, defined by simple differential
equations, which are in the one-to-one correspondence with
conductance distributions for quasi-one-dimensional systems
of size $L^{d-1}\times L_z$, characterizing by parameters
$L/\xi$ and $L_z/L$ ($\xi$ is the correlation length,
$d$ is the dimension of space). This
family contains the Gaussian and log-normal distributions,
typical for the metallic and localized phases.
For a certain choice of parameters, we reproduce the results for
the cumulants of conductance in the space dimension
$d=2+\epsilon$ obtained in the framework of the $\sigma$-model
approach.  The universal property of distributions is existence
of two asymptotic regimes, log-normal for small $g$ and
exponential for large $g$.  In the metallic phase they refer to
remote tails, in the critical region they determine practically
all distribution, in the localized phase the former asymptotics
forces out the latter.  A singularity at $g=1$, discovered in
 numerical experiments, is admissible in the framework of their
calculational scheme, but related with a deficient definition
of conductance.  Apart of this singularity, the critical
distribution for $d=3$ is well described  by the present theory.
One-parameter scaling for the whole distribution takes place
under condition, that two independent parameters characterizing
this distribution are functions of the ratio $L/\xi$.
}
 }

\date{}
\maketitle

\textwidth 6.4 in
\textheight 8.5 in

%\begin{document}
\setcounter{footnote}{0}
\setcounter{equation}{0}
\setcounter{figure}{0}
\setcounter{table}{0}
%\vspace*{2mm}

\begin{center}
{\bf 1. Introduction}
\end{center}

Dimensionless conductance $g=hG/e^2$ is determined
by conductance $G=\sigma L^{d-2}$ of a system in quantum
units $e^2/h$; the system is supposed to have a form of
the $d$-dimensional cube with a side $L$, and $\sigma$
is conductivity. One of the fundamental problems in theory of
disordered systems is related with the conductance
distribution $W(g)$ \cite{1}--\cite{20}. Its actuality was
realized after discovery of the
so called "universal conductance
fluctuations" in the metallic state \cite{1,2}
$$
\left\langle (\delta g)^2 \right\rangle\,=\,c \sim 1 \,,
\eqno(1)
$$
where the constant $c$ in the right-hand-side does not depend
on the system size and the strength of disorder, but depends
on the dimension of space and
the boundary conditions. Near the Anderson
transition the average value  $\langle g \rangle$ is also of
the order of unity, so conductance is a strongly fluctuating
quantity and may be not adequately  described by its first moment.
Investigation of higher moments in the framework of the
sigma-model approach \cite{3,4} leads
to the following results for
the cumulants of conductance\,\footnote{\,Recall that the
characteristic function $F(t)=\left\langle e^{igt} \right\rangle$
is the generating function for moments
(\,$F(t)=\sum_{n=0}^{\infty}(it)^n\left\langle g^n
\right\rangle/n!$\,), while its logarithm is
the generating function of cumulants
(\,$\ln F(t)=\sum_{n=0}^{\infty}(it)^n\left\langle\!
\left\langle g^n \right\rangle\!\right\rangle/n!$\,). In
particular, the second cumulant is a dispersion of the
distribution.   }  at
the transition point for the space dimension $d=2+\epsilon$
$$
\left\langle\!\left\langle g^n \right\rangle\!\right\rangle
\sim \left \{ \begin{array}{cc}
\,\,\epsilon^{n-2}\,,& n<n_0 \\
%{    }\\
\,\,L^{\epsilon n^2-2n}\,, & n>n_0 \end{array}
\right.\,,\eqno(2)
$$
where $n_0\sim 1/\epsilon$. The system size  $L$ in the second
relation is dimensionalized by a microscopic scale like the
mean-free-path $l$ or the lattice spacing $a$.
It led the authors of \cite{3,4} to conclusion
%It was a reason to conclude \cite{3,4}
on violation of the one-parameter scaling hypothesis
\cite{5}, according to which $g$ is completely determined by the
ratio $L/\xi$, where $\xi$ is the correlation length.

This conclusion was contested by Shapiro \cite{6,7,8},
who argued that divergency of high moments may be
determined by a negligible part of the distribution in its far
tail, while the main part of the distribution may obey
one-parameter scaling\,\footnote{\,Analogous arguments
were put forward in  \cite{4a}, where solution
of a certain hierarchical model led to the distribution
$W(g)$ with the  power-law tail obeying  one-parameter scaling.
}.  Using the approximate Migdal--Kadanov scaling transformation,
Shapiro has obtained for the distribution  $P(\rho)$ of
dimensionless resistances  ($\rho=1/g$) at the critical point
$$
P_c(\rho) = {\rm const}\, (\rho+1)^{-\alpha} \,,
\eqno(3)
$$
where $\alpha=1/\epsilon$ for $d=2+\epsilon$  \cite{6,7}.
If the analogous result was obtained for the conductance
distribution $W(g)$, it would give explanation of (2):
the  moments with  $n\agt 1/\epsilon$ do not exist in the infinite
system, and diverge with $L$ in a finite one.  Using the
requrrent relations, analogous to those for moments of $P(\rho)$,
Shapiro has constructed an example of the conductance distribution
$W(g)$, possessing the properties (2) and well defined in the
thermodynamic limit; in  analogy with (3) it has the power-law
behavior  $W(g)\sim g^{-2/\epsilon}$ for large  $g$. In spite of
the evident success, the latter result is questionable:
extrapolation to $\epsilon \sim 1$ gives a power-law tail
with the exponent of the order of unity,
contradicting to all numerical experiments \cite{10}.

The latter contradiction looks rather fundamental, since
at first glance  Eqs.2 unambiguously indicate
the power-law dependence with the exponent of the order
 $1/\epsilon$. This puzzle is resolved in the present paper.
We show that results (2) are not necessarily related with
the power-law tail of the critical distribution and compatible
with its exponential behavior at infinity; in this case, the
first result (2) is valid for all cumulants. As for the
second result (2), it corresponds to the situation when the
critical distribution is subjected to perturbation $\delta W(g)$,
retaining the system in the critical state\,\footnote{\,In the
renormalization group language \cite{12,13}, it means that the
system is deviated from the fixed point, but remains at the
critical surface. }. Due to the normalization condition,
a perturbation $\delta W(g)$ is necessarily alternating and
its relaxation is described by the diffusion type equation.
With the increase of the system size, $\delta W(g)$ undergoes a
diffusion spreading and tends to zero in the limit $L\to\infty$;
however, its tail extends to infinity and provides a divergency
of high moments. The strange analogy between $P(\rho)$ and
$W(g)$ discovered by Shapiro is also explained: these
distributions are described by equations of the same structure in
the space dimension $d=2+\epsilon$.

The main source of information on $W(g)$ are numerical
experiments. At present, there is a common belief that
the conductance distribution is Gaussian in the metallic state and
log-normal in the localized phase, while at the critical
point it is close to one-sided log-normal  \cite{9}
(Fig.1). More detailed investigation \cite{10} shows (Fig.2),
\begin{figure*}
\centerline{\includegraphics[width=6.2 in]{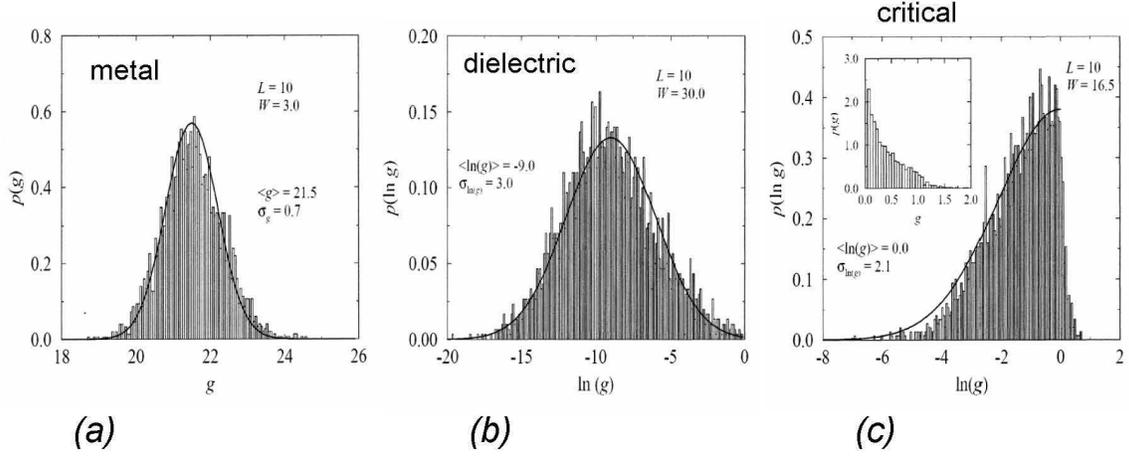}} \caption{
The conductance distribution is Gaussian in the metallic regime
(a), log-normal in the localized phase (b), and close
to one-sided log-normal at the critical point (c)
(according to \cite{9}). Solid lines correspond to the Gaussian
law.  }
\label{fig1} \end{figure*}
that the critical distribution can be divided into
two parts:  on the right of a certain point $A$ the
logarithm of $W(g)$ is linear in  $g$ (Fig.2,a), while on the
left of $A$ the logarithm of $P(\ln g)$ is quadratic in
$\ln g$ (Fig.2,b).  According to Markos \cite{10},  point $A$
is a real singularity, which was confirmed by Muttalib et al
\cite{11} in the framework of a certain theoretical scheme.
However, according to the general principles of the
modern theory of critical phenomena \cite{12,13}, singularities
are absent in finite systems and may arise only in the
thermodynamic limit; it contradicts to the accepted stationarity
of the critical distribution, which is formed at sufficiently
small $L$ and then remains unchanged. Below we reproduce
all listed properties and discuss the problem of a singularity.

\begin{figure*}
\centerline{\includegraphics[width=5.6 in]{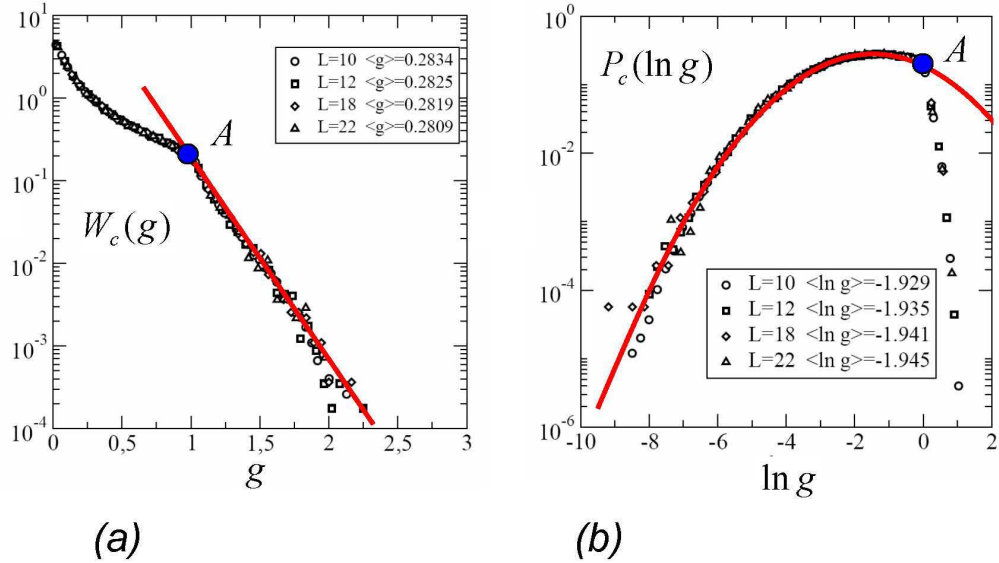}} \caption{
The distribution of conductance $W(g)$ (a) and
of the conductance logarithm $P(\ln g)$
($W(g)dg\equiv P(\ln g) d\ln g$) (b) according to numerical
data by Markos  \cite{10}.
On the right of point $A$ the logarithm of $W(g)$ is linear in
$g$, and on the left of $A$ the logarithm of $P(\ln g)$
is quadratic in $\ln g$.  } \label{fig2}
\end{figure*}

Analysis of the present paper is based on a modification of
the Shapiro approach \cite{6,7,8}. We introduce the two-parameter
family of distributions, defined by simple differential
equations, which are in  the one-to-one correspondence with
conductance distributions for
quasi-one-dimensional systems of size $L^{d-1}\times L_z$,
characterizing by parameters  $L/\xi$ and $L_z/L$. This family
contains the Gaussian and log-normal distributions, typical
for the metallic and localized phases. For a certain choice of
parameters, we reproduce results (2) for the cumulants in the space
dimension $d=2+\epsilon$. The universal property of distributions
is existence of two asymptotic regimes, log-normal for small
$g$ and exponential for large $g$, while their actuality
depends on the specific situation. In the metallic phase,
a distribution is determined by the central Gaussian peak,
while two asymptotic regimes refer to its far tails. In the
critical region, the log-normal behavior is extended to a
vicinity of the maximum, as we have seen in Fig.2, and
practically all distribution is determined by two asymptotes. In
proceeding to the localized phase, the log-normal behavior
extends even more and forces out the exponential asymptotics to
the region of the remote tail. A singularity at  point $A$ is
admissible in the framework of the calculational scheme used in
\cite{10}, but related with a deficient definition of
conductance; it will be smeared out for the correct definition.
Apart of the latter moment, numerical data in Fig.2,b are
well described by the present theory.

According to numerical experiments, one-parameter scaling
is valid for the distribution $W(g)$ in whole. It is
established by investigation of percentiles  \cite{14}, and
by analysis of average quantities $\left\langle g \right\rangle$,
$\left\langle \rho \right\rangle$, $\left\langle \ln g
\right\rangle$ \cite{15}. The present analysis is based on
the assumption of one-parameter scaling and its results agree with
this assumption. For a final decision on existence of scaling for
the whole distribution, one should prove that two independent
parameters, characterizing this distribution,  (e.g.,
$\left\langle  g \right\rangle$ and $\left\langle (\delta g)^2
\right\rangle$) are functions only of the ratio $L/\xi$.
For the first parameter, this property was established in
\cite{23}  using the self-consistent theory of localization
\cite{100}; its validity for the second parameter
also looks very probable.

\begin{center}
{\bf 2. The Shapiro approach}
\end{center}

\begin{figure*}
\centerline{\includegraphics[width=6.2 in]{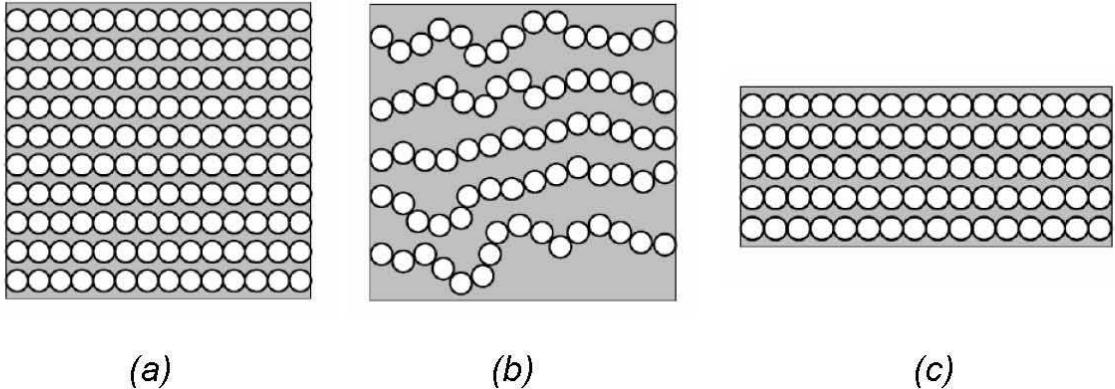}} \caption{
(a) In the Shapiro model, the $d$-dimensional system is
composed of $b^{d-1}$ one-dimensional chains, consisting of
$b$ atoms, inserted in the dielectric medium.  (b) A more
adequate interpretation corresponds to the change of the
artificial construction of Fig.3,a by a notion of the
resonant channels. (c) The "straighten" version of Fig.3,b.  }
\label{fig3} \end{figure*}

According to the Shapiro model \cite{6,7}, the $d$-dimensional
system is composed of one-dimensional chains, which are
considered to be independent of each other, i.e. separated by
dielectric interlayers (Fig.3,a). A distribution of
resistances $P_L(\rho)$ for each chain is described by
the equation
$$
\frac{\partial P_L(\rho)}{\partial L} =\alpha
\frac{\partial }{\partial \rho}
\left[\,(\rho^2+\rho) \frac{\partial P_L(\rho)}{\partial \rho}
\,\right]\,,
\eqno(4)
$$
which was derived by various authors \cite{17}--\cite{20}
and is considered to be sufficiently universal. The parameter
$\alpha$ is proportional to the dispersion of a random
potential and defined by relation  $\langle \rho
\rangle =\alpha L$ for small $L$; it has a sense of the inverse
correlation length of the $1D$ system. To obtain a description of
the $d$-dimensional system composed of $n=b^{d-1}$
one-dimensional chains, one should introduce the distribution
$W(g)$, corresponding to $P(\rho)$,
$$
W(g) =\int P(\rho) \delta\left(g\!-\!1/\rho \right)  d\rho
= g^{-2} P(1/g)
\eqno(5)
$$
and find a distribution of the sum of $n=b^{d-1}$ random
quantities with the same distribution $W(g)$. It can be made
by introducing the characteristic function
$F(t)=\left\langle e^{igt} \right\rangle$ and raising it to the
power $n$. The equation for $W(g)$, corresponding to (4),
is obtained by substitutions $P=g^2 W$, $\rho=1/g$:
$$
\frac{\partial W(g)}{\partial L} =\alpha
\left[\,2g(1+g) W(g) +g^2(1+g) W'_g(g)\,\right]'_g \,.
\eqno(6)
$$
Instead of the characteristic function, it will be more
convenient to use the Laplace transform
$$
F(\tau) = \int_0^\infty e^{-\tau g} W(g) dg \,,
\eqno(7)
$$
obtained by the change $it\to -\tau$. Multiplying (6) by
$e^{-\tau g}$ and integrating over $g$, one obtains the
equation for $F(\tau)$ corresponding to the $1D$ chain
$$
F_{L+\Delta L}(\tau) = F_{L}(\tau) + \alpha \Delta L
\left[\,-\tau^2 F'''_{L}(\tau) +  \right.
$$
$$
\left.
+\tau(\tau-1) F''_{L}(\tau)
\right] \,,
\eqno(8)
$$
which we have written for finite increments. Raising
$F_{L}(\tau)$ to the power  $n=b^{d-1}$ and setting
$b=1+\Delta L/L$, we have an additional term
$(\Delta L/L)(d\!-\!1) F_L \ln F_L$ in Eq.8, and finally
\cite{7}
$$
\frac{\partial F(\tau)}{\partial \ln L} = \alpha L
\left[\,-\tau^2 F'''(\tau) +\tau(\tau-1) F''(\tau)+
\right.
$$
$$
\left.
+p F(\tau) \ln F(\tau) \right] \,,
\eqno(9)
$$
where  $p=(d\!-\!1)/\alpha L$. The quantity $\alpha L$
has a sense of $L/\xi$, and evolution in $L$ for fixed
 $L/\xi$ leads to a stationary distribution, corresponding to
the large length scales. Equation (9) describes the
transient process, when $L$ is increasing from  the atomic
scale $a$ to scales of the order $\xi$.  Eq.9
gives the adequate description for the model of Fig.3,a, but
in fact
was not investigated in Shapiro's papers
\cite{6,7}.  Instead,  the simplified scheme was
elaborated (Sec.5), where all $b^{d-1}$ chains were taken to be
identical.  This scheme did not allow to obtain the correct
results for the metallic phase, and the whole approach was
admitted unpromising by the author himself.

\begin{center}
{\bf 3. Modification of the method}
\end{center}

In fact, after a certain modification the Shapiro
scheme becomes very fruitful. First of all, let us
change an interpretation of the model, coming from the artificial
construction of Fig.3,a to a more adequate version presented
in Fig.3,b. It is commonly accepted \cite{21}, that in the
strong disorder regime
conductance   is determined by the
resonant channels. There is a finite probability, that along a
properly chosen trajectory the fluctuations of the random
potential will be essentially less that in average over the
system. An increased (in comparison with $L$) length of
the trajectory  is compensated by a more essential diminishing of
$\alpha$, so
%in the localized regime
conductance $g\sim \exp(-\alpha L)$ of the resonant
channel  will be exponentially greater
than conductance of a typical chain in the Shapiro model
(Fig.3,a)\,\footnote{\,For illustration, consider
an example of the resonant trajectory on the $d$-dimensional
cubical lattice. Let the trajectory starts on the left side
of the system and is constructed
%in accordance with
by the following algorithm. If the trajectory comes to a certain
point $A$, then it can be
continued along $2d$ directions to the nearest neighbors.
%Of these continuations, we exclude directions to the left and
%backward along the trajectory; of other continuations we
We exclude of these continuations the directions to the left
and backward along the trajectory; as for the rest of
continuations, we choose the direction
to the site with the minimal value of the
random potential. It is easy to understand that (for large $d$)
the length of the trajectory will be $\sim Ld$, while the
amplitude $W$ of the random potential along it will be
approximately $d$ times less than in average. Since
$\alpha\sim W^2$, then a value of $\alpha L$ will be
approximately $d$ times smaller than for a typical straight-line
trajectory.  }.  Thus, we naturally come to the notion of $1D$
chains inserted in the dielectric medium. In this interpretation
we easily remove the internal inconsistency of the
initial model; indeed, a parallel connection of $L^{d-1}$ chains
with conductance $\exp(-\alpha L)$ gives
$$
g\sim L^{d-1}\exp(-\alpha L)\,,
\eqno(10)
$$
which tends to zero in the large $L$ limit. In fact,
statistics of the resonant channels essentially depends on $L$,
and the analogous dependence arises for the
parameter $\alpha$, making the zero limit in
(10) to be not obligatory.\,\footnote{\,The $L$ dependence of
$\alpha$ was introduced by Shapiro on the physical
grounds, but there is no reasons for it in the model of Fig.3,a.}

Since the typical length of the resonant trajectories is somewhat
greater than $L$, and their number is proportional to the
cross-section area but somewhat less than $b^{d-1}$, so
the "straighten" version of Fig.3,b corresponds to Fig.3,c,
i.e. a cube is transformed into a parallelepiped. If the sides
of the parallelepiped have the same scaling in $L$, then the
system is topologically $d$-dimensional and its conductance
distribution $W(g)$ possesses all qualitative properties of the
$d$-dimensional distribution. Nevertheless, there is a
quantitative difference related with transformation of the cube
into the parallelepiped, which is an uncontrollable effect in the
Shapiro scheme. The magnitude of this effect depends on the
specific situation.  Indeed, for weak disorder we practically
return to the initial model (Fig.3,a), since conductances of $1D$
chains fluctuate weakly and the current streamlets are
practically parallel to transversal boundaries of the system; so
the dielectric interlayers have practically no effect.
\begin{figure}
\centerline{\includegraphics[width=3.0 in]{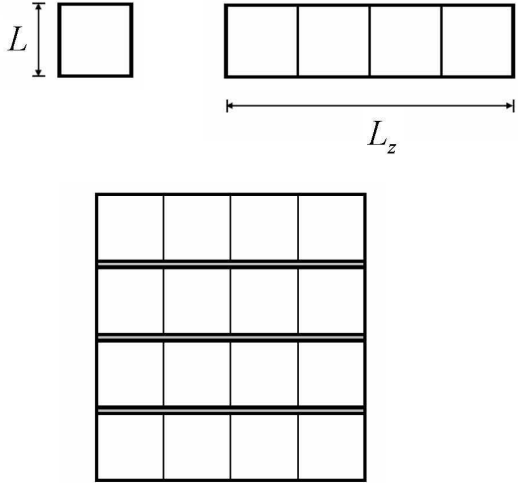}} \caption{
Large-scale constructions implied in the Shapiro scheme:
the cubes of size $L$ are arranged into quasi-one-dimensional
systems of length $L_z$, whose parallel connection composes
the $d$-dimensional system.  }
\label{fig4}
\end{figure}

Analogous conclusions can be drawn, if a situation is considered
for the large length scales. In fact, the Shapiro approach implies
the large-scale constructions: firstly, $b$ cubes are connected
successively to form a quasi-one-dimensional system  (Fig.4),
and then a parallel connection of  $b^{d-1}$
quasi-1D chains composes the $d$-dimensional system. For
large $L$, a concentration of the auxiliary dielectric phase
isolating quasi-1D chains becomes smaller and its influence
reduces, suggesting validity of the Shapiro scheme in the
large-scale limit. However, it was assumed implicitly that
properties of  quasi-1D systems are the same as those of
strictly one-dimensional chains. This assumption is partially
correct but needs an additional argumentation. It is commonly
accepted that all properties of the cubic system of size $L$ are
completely determined by the ratio $L/\xi$, so
$$
g=F(L/\xi) \,,
\eqno(11)
$$
in accordance with the one-parameter scaling hypothesis
\cite{5}\,\footnote{\,It would be more correctly to write that
the distribution $W(g)$ is determined by the ratio $L/\xi$,
but we prefer a simplified notation.}. If cubes are connected
to form a quasi-1D system, then its conductance depends on the
properties of the single block ($L/\xi$) and a numbers of cubes
($b=L_z/L$), i.e.
$$
g=F(L/\xi, L_z/L) \,.
\eqno(12)
$$
It is easy to verify that the distribution of $g$ depends
essentially on each of two parameters, and not on a certain
combination of them. Indeed, for the first two moments one has in
the metallic phase
$$
\left\langle g \right\rangle = \sigma L^{d-1} L_z^{-1}=
\left(L/\xi \right)^{d-2} \left(L/L_z \right)\,,
$$
$$
\left\langle (\delta g)^2 \right\rangle = f(L_z/L) \,,
\eqno(13)
$$
where the correlation length $\xi$ is introduced in the usual
manner \cite{101}. The latter result follows from the theory
of universal conductance fluctuations \cite{1,2}:
since the constant $c$ in (1) depends on the space
dimension, then the  function $f(x)$ is equal  $c_{d-1}$ for
$x\ll 1$, comes through $c_d$ at $x=1$  and tends to $c_1$
for $x\gg 1$.  Setting $L=a$ in Eq.12, we come to
conclusion that a conductance distribution for a quasi-1D
system corresponds to a certain  distribution
of a strictly 1D system\,\footnote{\,This statement is
rigorous in the framework of orthodox scaling suggested in
\cite{5}. In fact, the universal functions of type (12)
arise only at large length scales, while at scales
$\sim a$  they have a certain transient behavior. As was
discussed by Wilson \cite{12}, such transient behavior can be
excluded, if a model at small scales is chosen in the special
manner (see the "ideal RG trajectory" in \cite{102},
which is approximately realized in the so called
"improved" models \cite{103}). Since for 1D systems we  use
the equation possessing the high level of universality (see
below), the indicated "ideal" model will be also described by
this equation.  }.  At this point we discover the real defect of
the original Shapiro scheme: according to Eq.4, $P_L(\rho)$ is
determined by one parameter  $\alpha L$, and not two, as it
follows from Eq.12.

In fact, equation (4) allows two-parameter
generalization
$$
\frac{\partial P_L(\rho)}{\partial L} = \tilde\alpha\,
\left[\,-\gamma(2\rho+1) P_L(\rho) +
(\rho^2+\rho)
P'_L(\rho) \,\right]'
\eqno(14)
$$
(primes correspond to derivatives over $\rho$), which has
practically the same level of universality. Indeed, description
of 1D systems is conveniently made, considering each scatterer
as a "black box", characterizing by a transfer-matrix $\hat T$,
connecting amplitudes of  plane waves on the left
($Ae^{ikx}+Be^{-ikx}$) and on the right ($Ce^{ikx}+De^{-ikx}$)
of the scatterer (Fig.5,a):
$$
\left ( \begin{array}{cc} C \\ D \end{array} \right)
= \hat T
\left ( \begin{array}{cc} A \\ B \end{array} \right)\,.
\eqno(15)
$$
If the scatterers are arranged successively (Fig.5,b),
their transfer-matrices are multiplied. The matrix $\hat T$
is determined by the amplitudes of transmission ($t$) and
reflection ($r$) and in the presence of time reversal
invariance can be parametrized in the form \cite{24}
$$
\hat T= \left ( \begin{array}{cc} 1/t^* & - r^*/t^* \\
- r/t & 1/t \end{array} \right)\,=
$$
$$
= \left ( \begin{array}{cc} \sqrt{\rho\!+\!1}\, e^{-i\varphi} &
-\sqrt{\rho} \,e^{-i\theta}
\\ -\sqrt{\rho}\, e^{i\theta} & \sqrt{\rho\!+\!1}\, e^{i\varphi}
\end{array} \right)\,,
\eqno(16)
$$
where $\rho=|r/t|^2$ is the Landauer resistance \cite{25}. For
the product of large number $n$ of transfer-matrices,
the distribution of $\varphi$ and $\theta$ is usually stabilized,
i.e.
$$
P_n\left(\rho, \varphi, \theta\right)=
P_n\left(\rho\right) P\left(\varphi, \theta\right)\,.
\eqno(17)
$$
If the phase distribution is uniform  ($P\left(\varphi,
\theta\right)=const$), then we come to Eq.4,
while in the general case we obtain Eq.14 with
parameters (see Appendix 1)
$$
\gamma=\frac{1-2A_0}{2A_0},
\quad \tilde\alpha =2\alpha A_0,
\quad A_0= \left\langle \sin^2(\varphi-\theta) \right\rangle.
\eqno(18)
$$
It is easy to see  inequality $\gamma\ge
-1/2$, which will be essential for the phase diagram (Sec.6.5).
For standard microscopical models, Eq.14 with variable $\gamma$
arises for small $L$, when the distribution of $\varphi$ and
$\theta$ is not stabilized yet, and evolution of $\gamma$ is
not universal.  Equation (14) with constant $\gamma$ arises in
lattice models due to effects of commensurability of the wave
vector $k$ with a lattice spacing $a$ (when $ka$ is a rational
number) \cite{22}; it looks as a hardly observable exotic.
However, there exist systematical reasons for appearing of
equation (14), which are discussed in the next
section\,\footnote{\,In a somewhat different context, a necessity
of the two-parameter description of 1D systems was motivated in
paper  \cite{22a}. }.
\begin{figure}
\centerline{\includegraphics[width=3.4 in]{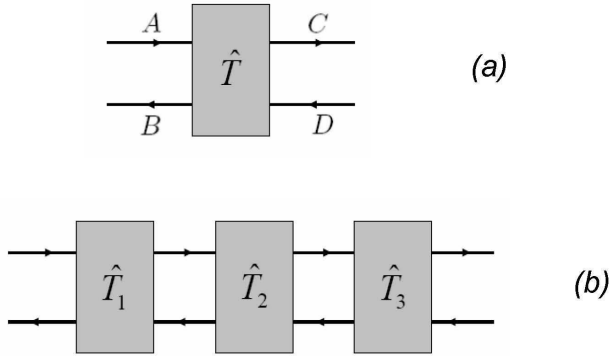}}
\caption{
(a) For 1D systems, the transfer-matrix $\hat T$ relates
amplitudes of plane waves on the left and on the right
of a scatterer.  (b) A successive arrangement of scatterers
corresponds to multiplication of transfer-matrices.
} \label{fig5} \end{figure}

Equation (14) is of the diffusion type with a "time"
$t=\tilde\alpha L$; so any initial distribution $P_0(\rho)$,
localized in the region of small $\rho$, at large times
transforms into the universal distribution, which has a different
form in the region of small and large $\rho$:
$$
P(\rho,t)=\frac{1}{\Gamma(\gamma\!+\!1)}\,
\frac{\rho^\gamma \exp\{-\rho/t\}}{t^{\gamma+1} }
\,, \qquad  \rho\alt 1
\eqno(19)
$$
$$
P(\rho,t)=\frac{1}{\rho \sqrt{4\pi t}}
\exp\left\{-\frac{[\ln \rho-(2\gamma+1)t]^2}{4t}\right\}, \quad
\rho\agt 1
\eqno(20)
$$
These results are  obtained, if the main in $\rho$ terms
are retained before $P_L(\rho)$ and $P'_L(\rho)$  in Eq.14. The
first distribution is close to Gaussian for large $\gamma$
$$
P(\rho,t)=\frac{1}{\sqrt{2\pi \gamma t^2}}
\exp\left\{-\frac{( \rho-\gamma t)^2}{2\gamma t^2}\right\}
\eqno(21)
$$
and arises at large times when the diffusive spreading
exceeds the width of the initial  distribution
$P_0(\rho)$.  In the opposite case one can neglect the
second term in the square brackets of (14) and obtain the
auto-model solution
$$
P(\rho,t)=e^{-2 \gamma t}
P_0\left\{\left( \rho+\textstyle{\frac{1}{2}} \right)
e^{-2 \gamma t} - \textstyle{\frac{1}{2}}\right\}\,,
\eqno(22)
$$
reducing to a pure drift for $\gamma t\ll 1$
$$
P(\rho,t)=P_0\left( \rho- \gamma t\right)\,.
\eqno(23)
$$
If the initial distribution is Gaussian, then Eqs.21,23
correspond to the commonly accepted view on the conductance
distribution in the metallic regime; this result was a
stumbling block in the original version of the Shapiro
method. Eq.20 corresponds to the
log-normal distribution, which is commonly accepted
for the localized regime.
\vspace{2mm}

\begin{center}
{\bf 4. Semi-transparent boundaries}
\end{center}

Is was argued in \cite{23}, that for a correct definition of
conductance of finite systems it is convenient to introduce
semi-transparent boundaries separated the system from the ideal
leads attached to it (Fig.6,a). The ideal leads are supposed to
be sufficiently massive which is necessary for the correct
interpretation of the linear response formulas. Indeed, the
latter demand that the entering them $\delta$-functions were
spread to the width $\Gamma$, which is tending to zero only
after transition to the thermodynamic limit. In the case of
finite systems, the thermodynamic limit is realized by
%due to
increasing the size of ideal leads  \cite{26}. Such
definition of conductance refers to the composite system
"sample+ideal leads" and
a question arises
on its relation to the system under consideration.

\begin{figure}
\centerline{\includegraphics[width=3.0 in]{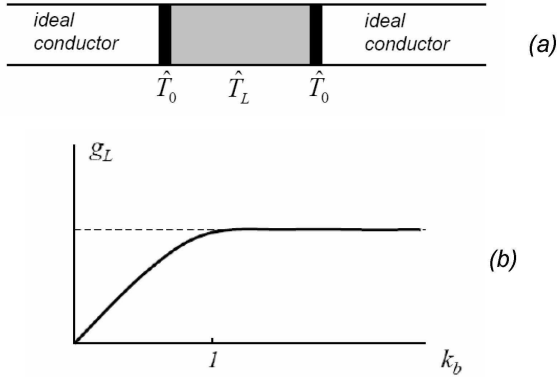}} \caption{
(a) For a correct definition of  conductance of finite
systems, it is useful to introduce semi-transparent interfaces,
separated the system from the ideal leads connected to it. (b)
Conductance of a finite system as a function of the parameter
$k_b$, having a sense of the effective transparency
of the boundary.  } \label{fig6}
\end{figure}
% %

If $k_b$ is a parameter, characterizing the effective
transparency of  interfaces, then the $k_b$ dependence
of conductance has a view shown in Fig.6,b \cite{23}:
the linear behavior valid for small $k_b$ (when resistance
is determined by weakly-transparent interfaces), is changed by
saturation at  $k_b\sim 1$, when the spreading of discrete levels
of a finite system becomes comparable with their spacing,
and a continuous density of states is formed. The
physically reasonable definition of conductance corresponds
to a value at the plateau for  $k_b\agt 1$; however, one should
not take too large $k_b$, because the "plateau" may
correspond to a slow dependence arisen due to
influence on the system of its environment. Instead of setting
$k_b\sim 1$ in the dependence $ g=f(k_b)$ one can take a
derivative $d g /dk_b$ for $k_b\to 0$;\,\footnote{\,These
two procedures coincide only
in the order of magnitude, but it should not arouse anxiety:
only a ratio of conductances has a physical sense and
variation of the definition by a constant factor
corresponds to a change of the unit of measurement.}
then conductance of the open system is defined in terms of almost
closed systems \cite{23}.  Such definition (a) surely refers to
the system under consideration and does not depend on the
properties of the environment, (b) is free from ambiguities
related with exclusion of the contact resistance of the reservoir
\cite{27}, and (c) provides infinite conductance for
an ideal system \cite{23}. It is natural to expect that
introducing of semi-transparent boundaries will be also
useful in discussion of conductance distributions; but it
immediately leads to equation (14) with finite  $\gamma$.

If $\hat T_L$ is the transfer-matrix for a system of size $L$,
then a situation corresponding to Fig.6,a is described by the
matrix
$$
\hat T'_L = \hat T_0 \hat T_L \hat T_0\,,
\eqno(24)
$$
where $\hat T_0$ corresponds to the interface. The increase
of  the system length by  $\Delta L$ gives
$$
\hat T'_{L+\Delta L} =
\hat T_0 \hat T_L \hat T_{\Delta L} \hat T_0=
\hat T'_L \cdot \hat T_0^{-1} \hat T_{\Delta L} \hat T_0\,,
\eqno(25)
$$
and corresponds to multiplication by the matrix
$\hat T_0^{-1} \hat T_{\Delta L} \hat T_0$ close to the unit one,
so the evolution equation can be derived by a standard method
(see Appendix 1). If $\hat T_{L}$ is accepted in the form (16)
and  $\hat T_{0}$  corresponds
to the point scatterer
$$
\hat T_0= \left ( \begin{array}{cc} 1-i\kappa &  -i\kappa
\\ i\kappa & 1+i\kappa \end{array} \right)\,,
\eqno(26)
$$
then for $\kappa\gg 1$ one has a relation between parameters of
$\hat T'_{L}$ and $\hat T_{L}$:
$$
\rho'= 4\kappa^4 \left ( \sqrt{\rho\!+\!1}
\sin{\varphi}+ \sqrt{\rho}\sin{\theta} \right)^2\,,
\eqno(27)
$$
$$
{\rm tg}\varphi'= 2\kappa
\frac{\sqrt{\rho\!+\!1}
(\cos{\varphi}-\kappa\sin{\varphi})-\kappa\sqrt{\rho}
\sin{\theta} } {\sqrt{\rho\!+\!1}
(\cos{\varphi}-2\kappa\sin{\varphi})-2\kappa\sqrt{\rho}
\sin{\theta} }\,, \eqno(28)
$$
$$
{\rm tg}\theta'= -2\kappa
\frac{\sqrt{\rho\!+\!1}
(\cos{\varphi}-\kappa\sin{\varphi})-\kappa\sqrt{\rho}
\sin{\theta} } {\sqrt{\rho} \cos{\varphi} } \,.
\eqno(29)
$$
It is easy to see that the distribution of $\varphi'$ and
$\theta'$ is not trivial even for the uniform phase
distribution $P(\varphi,\theta)$ of the initial matrix $\hat
T_L$.  In particular, for $\kappa\gg 1$ distributions of
$\varphi'$ and $\theta'$ have a form of the narrow Lorentz
peaks
$$
P(\varphi')= \frac{1}{2\pi} \frac{1/2\kappa^2}
{1/4\kappa^4+(\delta \varphi'+1/\kappa)^2} \,,
\eqno(30)
$$
$$
P(\theta')= \frac{1}{2\pi} \frac{1/2\kappa^2}
{1/4\kappa^4+(\delta \theta')^2} \,,
\eqno(31)
$$
where $\delta\varphi'=\varphi'\pm \pi/2$ and
$\delta\theta'=\theta'\pm \pi/2$. We have taken into account
that for  $\rho\gg 1$ quantities  $\varphi'$ and $\theta'$
does not depend on $\rho$; a certain dependence on $\rho$
is present for $\rho\alt 1$, but localization at
$\delta\varphi'\approx 1/\kappa$ and
$\delta\theta'\approx 0$ is  retained as one can
immediately see from (28), (29). Calculating the average value
in (18) using distributions (30), (31), one has
$$
A_0=1/\kappa^2\,, \qquad\gamma= \kappa^2/2 \,.
\eqno(32)
$$
We come to the following conclusion: if the phase distribution
$P(\varphi,\theta)$ for the initial system is uniform, then
introducing of weakly-transparent interfaces makes it
strongly localized  and leads to the evolution equation with
$\gamma\gg 1$.

\begin{center}
{\bf 5. Simplified schemes}
\end{center}

In the complete version of the Shapiro approach, an evolution
of the distribution is described by a nonlinear equation for
$F(\tau)$, which is somewhat cumbersome for investigation.
There is a certain methodical interest in formulation of
approximate schemes leading to the more simple equations.

\begin{center}
{\small\bf 5.1. Shapiro's scheme for average quantities}
\end{center}

Multiplying (4) by $\rho$ and integrating, one has the closed
equation for the average resistance of
the 1D system, whose solution
$$
\bar\rho_L={\textstyle\frac{1}{2}} \left(e^{2\alpha L}-1 \right)
\eqno(33)
$$
coincides with known results \cite{7,24,25}. Composing a system
of length $bL$ by successive connection of $b$ blocks of
size $L$, one has a scale transformation for the $1D$ chain
$$
\bar\rho^{(1)}_{bL}={\textstyle\frac{1}{2}}
\left[\left(1+2\bar\rho_L^{(1)}\right)^b -1\right] \,,
\eqno(34)
$$
while the parallel connection of $b^{d-1}$ chains composes
the $d$-dimensional system:
$$
\bar\rho_{bL}={\textstyle\frac{1}{2}} b^{-(d-1)}
\left[\left(1+2\bar\rho_L\right)^b -1\right]\,.
\eqno(35)
$$
Taking $b$ close to unity, one has the differential equation,
which can be rewritten in terms of the variable $g_L=1/2\bar\rho_L$
\cite{7}
$$
\frac{d \ln g_L}{d \ln L} = d-1 -\left(1+g_L\right)
\ln\left(1+1/g_L\right) \equiv \beta(g_L)
\eqno(36)
$$
and has a form expected from  one-parameter scaling \cite{5}.
Equation (36) gives a qualitative description of the Anderson
transition and reproduces the correct result $\nu=1/\epsilon$
for the critical exponent $\nu$ of the correlation length
in the space of dimension $d=2+\epsilon$. The latter is not
surprising, since the only essential assumption is made in
proceeding from (34) to (35): for the parallel connection of
chains one should sum average conductances, i.e.  $\bar g_{bL}=
b^{d-1}\bar g^{\scriptscriptstyle\, (1)}_{bL}$ instead of the
exploited relation  $\bar\rho_{bL}= b^{-(d-1)}
\bar\rho^{\scriptscriptstyle\,(1)}_{bL}$. The latter is valid
approximately for a narrow distribution, which is the case
for  $d=2+\epsilon$.

Using equation (14) instead of (4) and taking into account
that $\tilde\alpha(\gamma+1)=\alpha$ according to (18),
one can see that dependence on $\gamma$ disappears and
the result (33) retains for $\bar\rho_L$; so Eqs.34--36
remain unchanged. Therefore, the modified Shapiro
scheme leads to the correct critical behavior of  $\xi$,
so far as
%in the extent
this behavior can be controlled.

\begin{center}
{\small\bf 5.2. Shapiro's scheme for  $P(\rho)$}
\end{center}

In a simplified variant of the Shapiro approach, all chains
in Fig.3,a are assumed to be identical, so $P_L(\rho)= b^{d-1}
P^{\scriptscriptstyle\,(1)}_L(b^{d-1}\rho)$ and instead of (9)
one has a linear equation for $P_L(\rho)$
$$
\frac{\partial P_L(\rho)}{\partial \ln L} = A
\left[\,-\gamma(2\rho+1) P_L(\rho) + \right.
$$
$$
\left. +\rho(\rho+1)
P'_L(\rho) +p\rho P_L(\rho) \,\right]' \,,
\eqno(37)
$$
where $A=\tilde\alpha L$. It has the stationary solution
$$
P_c(\rho)
= {\rm const}\, \frac{\rho^\gamma}{(\rho+1)^{p-\gamma}} \,,
\eqno(38)
$$
coinciding with (3) for $\gamma=0$. Accepting the result
$\bar\rho=\epsilon$ for  $d=2+\epsilon$, following from
(36), one has $p=(\gamma+1)/\epsilon$. In opposite to
Shapiro's result (3), the distribution (38) provides finite
values of moments  $\left\langle g^n \right\rangle$
for  $n\alt \gamma$. Setting  $\gamma\sim 1/\epsilon^2$
and calculating the Laplace transform (7) by the
saddle-point method, one has
$$
F(\tau) = \exp\left\{ \frac{1}{\epsilon^2  } f(\epsilon
\tau)\right\}  \,,
\eqno(39)
$$
where $f(x)$ has a regular expansion and provides correct
results for the cumulants determined by the first relation (2);
the second relation (2) is not reproduced in this scheme.

\begin{center}
{\small\bf 5.3. A simplified scheme for $d=2+\epsilon$}
\end{center}

A more adequate approximation can be formulated
having in
mind a situation for  $d=2+\epsilon$. According to (2),
the cumulants  $\left\langle\!\left\langle g^n
\right\rangle\!\right\rangle$ decrease fast with $n$,
so in the main order in $\epsilon$  one can set
$\ln F(\tau)\approx const\,\tau$.
Not difficult to
trace that it is equivalent to appearing the term
$\rho^2 P$ in the square bracket of equation for  $P(\rho)$
%%
%It can be traced, that it leads to the term $\rho^2 P$
%in the square bracket of equation for  $P(\rho)$
$$
\frac{\partial P(\rho)}{\partial \ln L} = A
\left[\,c\rho^2 P(\rho)-\gamma(2\rho+1) P(\rho) +  \right.
$$
$$
\left. +\rho(\rho+1)
P'(\rho)  \,\right]' \,.
\eqno(40)
$$
Till the present moment we used the Landauer definition
of resistance  $\rho=|r/t|^2$ \cite{25}. The alternative
is given by the Economou-Soukoulis definition
$\tilde\rho=|1/t|^2$ \cite{26}, so $\rho=\tilde\rho-1$.
Taking into account ambiguities, related with  exclusion
of the contact resistance of the reservoir \cite{27}, and
a change of the $\rho$ normalization
in the course of transition to
quasi-1D systems, we should generally make a change
$\rho\to \rho-\rho_0$, where  $\rho_0$ depends on details
of the definition. However, in the case of weakly-transparent
boundaries, a scale of $\rho$ is increased by a factor
$\kappa^4$  (see (27)), so the terms containing $\rho_0$ are
insignificant and can be omitted. Such universal equation
(where $\rho(\rho+1)$ is replaced by $\rho^2$),
obtained in the limit  $\kappa\to\infty$, can be extrapolated
into the region $\kappa\sim 1$: it exactly corresponds to a
procedure suggested in \cite{23}, when the dependence $g=const\,
k_b$, obtained in the limit $k_b\to 0$, is extrapolated to  value
$k_b=1$. Replacement $\rho\to \rho-\rho_0$  should be made
also in other terms of (40), where a situation is more
complicated due to  unknown behavior of parameters in the
course of the described procedure; in fact, the effect of the
change $\rho\to \rho-\rho_0$ can be removed by redefinition
of parameters $c$, $\gamma$ and the change of the general
$\rho$ scale.
As a result, we have the equation for $P(\rho)$
$$
\frac{\partial P(\rho)}{\partial t} =
\left[\,(c\rho^2 -2\gamma\rho-\gamma) P(\rho) + \rho^2
P'(\rho)  \,\right]' \,,
$$
$$
\qquad t=A\ln L   \,,
\eqno(41)
$$
which gives the equation  for $W(g)$ of the same
structure
$$
\frac{\partial W(g)}{\partial t} =
\left[\,\left(\gamma g^2 +2(\gamma\!+\!1)g-c\right) W(g) +
g^2 W'(g)  \,\right]',
\eqno(42)
$$
and explains the strange analogy between $P(\rho)$ and $W(g)$,
discovered by Shapiro. The stationary solution has a form
$$
W_c(g)= {\rm const}\, g^{-2(\gamma+1)} \exp\{-c/g-\gamma g\} \,,
\eqno(43)
$$
and provides a finiteness of all moments of conductance.
Calculating the Laplace transform by the saddle-point method
and setting $\gamma\sim1/\epsilon$, $c\sim1/\epsilon^3$,
we obtained the result of type (39), providing validity of the
first relation (2).

For evolution of moments one has from (42)
$$
\frac{\partial \langle g^n\rangle}{\partial t} =
cn\langle g^{n-1}\rangle+n(n\!-\!2\gamma\!-\!1)
\langle g^{n}\rangle
-\gamma n \langle g^{n+1}\rangle
\eqno(44)
$$
and deviations  $x_n=\langle g^n\rangle-\langle
g^n\rangle_c$ from the stationary values obey the same
equation. If the latter are proportional to
$\exp(\lambda t)$, then we have a three-diagonal matrix for
determination of the $\lambda$ spectrum. The eigenvalue
$\lambda_n$ is determined by the matrix of size  $n\times n$
and corresponds to nonzero deviations  $x_1$, $x_2$, $\ldots $,
$x_n$, obtained for the boundary conditions $x_0=x_{n+1}=0$.
The indicated matrix is not Hermitian and its eigenvalues are
complex; one can find by the quasi-classical
method, that
$$
{\rm Re}\, \lambda_n= \frac{n^2/2-(2\gamma+1)n}{\ln n}\,,
\qquad n\alt 1/\epsilon^2  \,.
\eqno(45)
$$
In the region of values  $n\sim 1/\epsilon$ the denominator
can be replaced by  $\ln(1/\epsilon)$, and one obtains for
evolution of moments
$$
\langle g^n \rangle \sim e^{\lambda_n
t}\sim L^{\tilde A \left[n^2-2(2\gamma+1)n\right]}  \,,
\eqno(46)
$$
where  $\tilde A=A/2\ln(1/\epsilon)$. To reproduce the second
relation in  (2) one should accept $\tilde A=\epsilon$,
$(2\gamma+1)\tilde A=1$, so $\gamma\sim 1/\epsilon$ in
agreement with the condition for validity of (39).  In the
described simplified scheme, the second result (2)
is reproduced with the logarithmic accuracy; in the complete
theory (see below) it will be obtained precisely.

\begin{center}
{\bf 6. Conductance distribution in the complete
theory}
\end{center}

\begin{center}
{\small\bf 6.1. The main equations}
\end{center}

It is clear from above considerations, that for 1D systems
we should accept equation (14) with replacement
$\rho\to\rho-\rho_0$. Then we should take a limit of
weakly-transparent interfaces to obtain the universal
equation, and  extrapolate it to transparency of the order of
unity; practically it reduces to replacements $\rho(\rho+1)\to
\rho^2$ and $\gamma(1-2\rho_0)\to \tau_0$. As a result, the
equation for $F(\tau)$ in the $d$-dimensional case has a form
$$
\frac{\partial F(\tau)}{\partial t} =
\tau(\tau\!+\!\tau_0) F''(\tau)-2\gamma\tau F'(\tau) +p
F(\tau) \ln F(\tau) \,,
$$
$$
 \qquad t=A\ln L \,,
\eqno(47)
$$
where the parameter  $\tau_0$ should be
positive to avoid singularities on the positive semi-axis;
it specifies the general scale of conductance, which is not
controlled in the theory. Introducing a variable $u$
by relation  $F(\tau)=\exp\{u(\tau)\}$, we have
$$
\frac{\partial u}{\partial t} =
\tau(\tau\!+\!\tau_0)\left[\, u''+u'^2\right]
-2\gamma\tau u' +p u \,.
\eqno(48)
$$
The stationary version of equation (48) is of the main interest,
since this equation describes the transient behavior to the
limit of large length scales  (see Footnote 7) for fixed values
of $L/\xi$ and $L_z/L$. As a result,  $\xi$ is increasing to
infinity, and all obtained distributions correspond to the
Anderson transition point,  differing in values of two indicated
parameters\,\footnote{\,"The critical distribution", discussed in
\cite{6,7,8} and other papers, corresponds to a situation
$L_z=L$, $L/\xi=0$. These two conditions determine the critical
values $p_c$ and $\gamma_c$ for two parameters entering (48).  We
do not try to calculate these parameters for any specific
situations but investigate all family of distributions in whole.
In the framework of one-parameter scaling, the values $p_c$ and
$\gamma_c$ should depend only on $d$ and the boundary
conditions.}.  A stationary configuration can be
%searched
sought
in the form of the regular expansion
$$
u(\tau) = \sum\limits_{n=1}^{\infty}\,B_n \tau^n\,,
\eqno(49)
$$
where a zero-order term is absent due to normalization of $W(g)$
(see (7)). Substituting to (48), one has for
the expansion coefficients
$$
(p-2\gamma)B_1 +\tau_0 B_1^2 +2\tau_0 B_2=0  \,,
$$
$$
(p -4\gamma)B_2 +4\tau_0 B_1 B_2 +6\tau_0 B_3+2B_2+B_1^2=0 \,,
\eqno(50)
$$
$$
(p-6\gamma)B_3 +4\tau_0 B_2^2 +6\tau_0 B_1 B_3 +12\tau_0 B_4+
$$
$$
+6B_3+4B_1 B_2=0 \,,
$$
and so on.

\begin{center}
{\small\bf 6.2. Reproducing the results for $d=2+\epsilon$}
\end{center}

Coefficients  $B_n$ are proportional to the cumulants
$\left\langle\!\left\langle g^n \right\rangle\!\right\rangle$
and for $d=2+\epsilon$ they obey the hierarchy
$B_1\gg B_2 \gg B_3 \gg \ldots$ (see (2)); so we can omit
the term with  $B_2$ in the first equation, the term with
$B_3$ in the second equation, etc., which corresponds to
neglecting the term $\tau\tau_0 u''$ in Eq.48. After it,
coefficients  $B_1$, $B_2$, $\ldots$ are determined uniquely
and for $p\gg 1$, $\gamma\ll p$ are estimated as
$B_n\sim p/\tau_0^n$. It suggests substitution
$u(\tau)=p f(\tau/\tau_0)$ and the main order in $p$
gives the equation for  $f(x)$
$$
x(x+1) f'^2 +f=0\,.
\eqno(51)
$$
Its solution leads to the result
$$
F(\tau) =\exp\left(-p\, {\rm Arsh}^2\sqrt{\tau/\tau_0} \right)
\,,
\eqno(52)
$$
and the necessary form (39)  is obtained for $\tau_0\sim 1/\epsilon$,
$p\sim 1/\epsilon^2$. Producing the inverse Laplace
transformation
$$
W(g) = \frac{\tau_0}{2\pi i} \int\limits_{-i\infty}^{i\infty} \,dx
\exp\left\{-p\, {\rm Arsh}^2\sqrt{x} +\tau_0 g x \right\} \,,
\eqno(53)
$$
and calculating the integral in the saddle-point approximation,
we have after setting $g_c=p/\tau_0$
$$
W(g) \,\sim\, \frac{1}{g}
\exp\left\{- \frac{p}{4} \left(\ln\frac{a(g)g_c}{g} \right)^2
 \right\} \,, \quad g\ll g_c
 \eqno(54a)
$$
$$
W(g) \,\sim \,\exp\left\{- \frac{3}{4}p \left(\frac{g-g_c}{g_c} \right)^2
 \right\} \,, \quad |g- g_c|\ll g_c
 \eqno(54b)
$$
$$
W(g)\, \sim \,
\left(\frac{g_c}{g} \right)^{3/2}
 \exp\left\{  -\tau_0 g
 \right\} \,, \quad  g\gg g_c
  \eqno(54c)
$$
where $a(g)$ is logarithmically varying
function.  Eqs.$54a$ and $54b$ can be formally united, if
$a(g)$ is accepted to tend to unity for $g\to g_c$. For
$p\sim 1$, a difference of $a(g)$ from a constant is practically
inessential, and the log-normal asymptotics $(54a)$,
obtained formally for small $g$, describes satisfactorily
a vicinity of the maximum $g_c$; together with $(54c)$ it
explains the situation demonstrated in Fig.2.

\begin{figure*}
\centerline{\includegraphics[width=6.8 in]{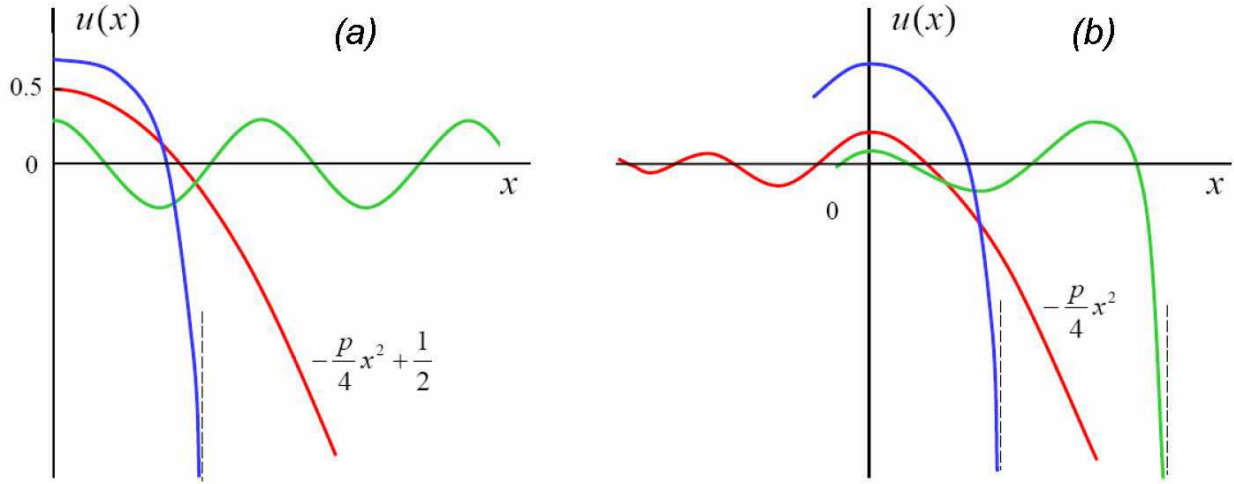}} \caption{
(a) Investigation of the quadrature (61) of Eq.60 with
$\tilde\gamma=0$ leads to existence of three types of
solutions: quadratic for $u(0)=1/2$, periodic for $u(0)<1/2$
and logarithmically divergent at a finite point for
$u(0)>1/2$.  (b) In the case $\tilde\gamma>0$, only the solution
with a quadratic asymptotics remains regular; all other
solutions, increasing from the region of negative $x$,
break to a logarithmic singularity, when their amplitude
of oscillations becomes of the order of unity.  }
\label{fig7} \end{figure*}

Consider evolution of cumulants, if the distribution is
deviated from the stationary one. Assuming the deviations
to be proportional to $\exp(\lambda t)$, we have the
equation
$$
\tau(\tau\!+\!\tau_0) u'^2 +pu= -\tau(\tau\!+\!\tau_0)
 u''
+2\gamma \tau u' +\lambda u \,,
\eqno(55)
$$
whose solution is known in the absence of the right-hand-side.
The latter can be taken into account iteratively, and
setting $u=u_c+\delta u$  one has
$$
\hat L \delta u = f\{u\} \,,
\eqno(56)
$$
where
$$
\hat L \delta u\equiv \tau(\tau\!+\!\tau_0)\,2 u_c'\delta u' +p\delta u\,,
$$
$$
f\{u\}\equiv -\tau^2 u'' +2\gamma \tau u' +\lambda u \,,
\eqno(57)
$$
and the  term $\tau\tau_0  u''$ is omitted, since it is
inessential for the actual solution (see beginning of Sec.6.2).
The operator $\hat L$ is proportional to  $p$, and substitution
$u=u_c$ into the right-hand-side of (56) gives a correction
$\delta u\sim u_c/p$ to the previously found solution.  Now, if
we introduce a perturbation of the form
$$
u=u_n(\tau)=B_n \tau^n\,, \qquad\lambda=\lambda_n=n^2 -(2\gamma+1)n
\, \eqno(58)
$$
in the right-hand-side of (56), then it is easy to see
that  $f\{u\}=0$ and $\delta u=0$. Thus,
in a framework of the iterative procedure
the perturbation (58) does not violate the validity of Eq.55
and can have a non-stationary evolution. Since coefficients
$B_n$ are proportional to cumulants  $\left\langle\!\left\langle g^n
\right\rangle\!\right\rangle$, then
$$
\left\langle\!\left\langle g^n \right\rangle\!\right\rangle
\sim e^{\lambda_n t} \sim L^{A[n^2 - (2\gamma+1)n]}
\eqno(59)
$$
and with a choice $A=\epsilon$, $(2\gamma+1)A=2$ we reproduce
the second relation (2).

It should be noted that our analysis is not restricted by
reproducing (2), but gives the closed expression (53) for the
critical distribution in the space dimension $d=2+\epsilon$.

\begin{center}
{\small\bf 6.3. Uniqueness of the physical
solution} \end{center}

We have seen above that in the main order in $p$
coefficients $B_n$ are determined unambiguously, providing
uniqueness of the stationary solution  $u_c(\tau)$.
If the omitted terms are taken into account iteratively,
and equation (56) is solved by variation of constants,
then $\delta u(\tau)=f_{reg}(\tau)+C{\rm
Arsh}\sqrt{\tau/\tau_0}$, where  $f_{reg}(\tau)$ is a regular
function. The correction to $u_c(\tau)$ is determined uniquely,
if $u(\tau)$ is accepted to be regular at zero. As we see
below, such situation retains in the general case.

For large  $\tau$, we can replace $\tau(\tau+\tau_0)$ by $\tau^2$
in equation (48), and substitution $x=\ln\tau$ reduces its
stationary version to the form
$$
 u_{xx}''+\left(u_{x}'\right)^2-\tilde\gamma u_x' +p u =0 \,,
\eqno(60)
$$
where  $\tilde\gamma=2\gamma+1$. For $\tilde\gamma=0$,
Eq.60 is integrated in quadratures \cite{28}
$$
x=C_1+\int \frac{du}{\sqrt{Y}}\, ,
%\quad\mbox{\rm where}
\quad Y=C_2 e^{-2u}+\frac{p}{2}\,(1-2u),
\eqno(61)
$$
and investigation of (61) leads to the picture presented in
Fig.7,a. Invariance relative to changes $x\to -x$ and $x\to
x+x_0$ allows to consider only solutions even in $x$, whose
derivative at $x=0$  is zero. If  $u(0)=1/2$, then Eq.60
has a simple solution
$$
u(x)= -\frac{p}{4}\,x^2 + \frac{1}{2}  \,;
\eqno(62)
$$
if $u(0)<1/2$, then solutions are periodical; if  $u(0)>1/2$,
then solutions diverge logarithmically at a finite point
$x_c$ (see Fig.7,a). In fact, all characteristic regimes
are determined by domination of two (from three) terms  in
Eq.60 with $\tilde\gamma=0$: domination of $u'^2$ and $pu$ leads
to the quadratic solution $u\sim x^2$, domination of $u''$ and
$pu$ provides the periodic solution $u\sim \cos{\sqrt{p}x}$,
domination of $u''$ and $u'^2$ results in the logarithmic
singularity $u\sim \ln(x-x_c)$.

In the case $\tilde\gamma>0$, invariance
respective $x\to -x$ is violated and the periodic solutions
acquire a negative damping decrement. As a result, the quadratic
solution (transforming to oscillations decreasing in the negative
$x$ direction) remains the only regular one. All other solutions
break to the logarithmic singularity after reaching the amplitude
of oscillations of the order of unity (Fig.7,b).

According to the definition (7), the function $F(\tau)$ is
regular and monotonically decreasing; so in the region
$x=\ln\tau\gg 1$ only the quadratic solution is physically
satisfactory; it has one-parameter indeterminacy related with
shifts along the $x$ axis. In the region $\tau\alt 1$ one should
return to the initial equation (48), and the one-parameter
freedom is removed by condition $u(0)=0$, following from
normalization of $W(g)$.

\begin{center}
{\small\bf 6.4. Universal tails} \end{center}

A typical  behavior of the function $u(\tau)$ is
represented in Fig.8. For large $\tau$ it has the asymptotics
$$
 u(x)=-{\textstyle \frac{1}{4}} p(\ln{\tau}-x_0)^2 \,,
\eqno(63)
$$

\noindent
which in the
saddle-point approximation gives the log-normal tail (54a)
in the small  $g$ region.

The asymptotics of a distribution for large $g$ is
determined by a singularity  at the point $\tau=-\tau_0$,
in whose vicinity the general solution has a form
$$
 u(\tau)= C_1(\tau+\tau_0)^{1+2\gamma} +C_2 \,.
\eqno(64)
$$
Substitution of (64) into the inverse Laplace transform gives
the exponential behavior for large $g$:
$$
W(g)\sim g^{-2-2\gamma} \exp\left( -\tau_0 g \right)
\,,\qquad g\to\infty \,.
\eqno(65)
$$
This result is more correct than (54c), where the character of
singularity at $\tau=-\tau_0$  was somewhat distorted due to
the use of approximation (52).

As a result, the tails of the distribution $W(g)$ are
universal, but their physical actuality depends on the
specific situation (see below).

\begin{center}
{\small\bf 6.5. Phase diagram\,\footnote{\,In this section
we set $\tau_0=1$. Results for arbitrary $\tau_0$
can be obtained by
the change  $\tau\to\tau/\tau_0$ or $g\to g\tau_0$ in final
expressions.  } } \end{center}

Calculating corrections to (52), including
the omitted terms  iteratively, one has
for the coefficients $B_1$ and $B_2$
$$
-B_1=p-2\gamma-{\textstyle \frac{2}{3}}\,,\qquad
3B_2=p-{\textstyle \frac{8}{3}} \gamma
-{\textstyle \frac{38}{45}} \,.
\eqno(66)
$$
Since $B_1=-\langle g\rangle$, $2B_2=\langle (\delta g)^2\rangle$,
then constant values of $\langle g\rangle$ and $\langle (\delta
g)^2\rangle$ correspond to straight lines in the $(p,\gamma)$
plane. Formally, equation (66) is valid for $p\gg 1$,
$\gamma\ll p$, but practically it describes almost all phase
diagram (Fig.9,a).

%%
%%%\vspace{5mm}
%%
\begin{figure}
\centerline{\includegraphics[width=2.9 in]{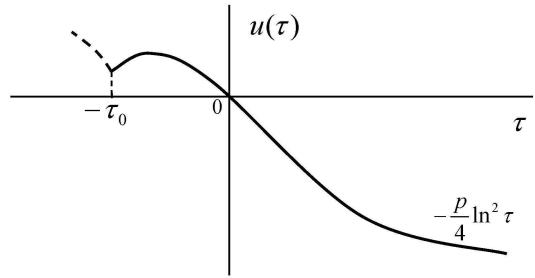}}
\caption{
Behavior of the physical solution $u(\tau)$ on the real axis;
for $\tau<-\tau_0$ it becomes complex.  } \label{fig8} \end{figure}
\begin{figure*}
\centerline{\includegraphics[width=4.5 in]{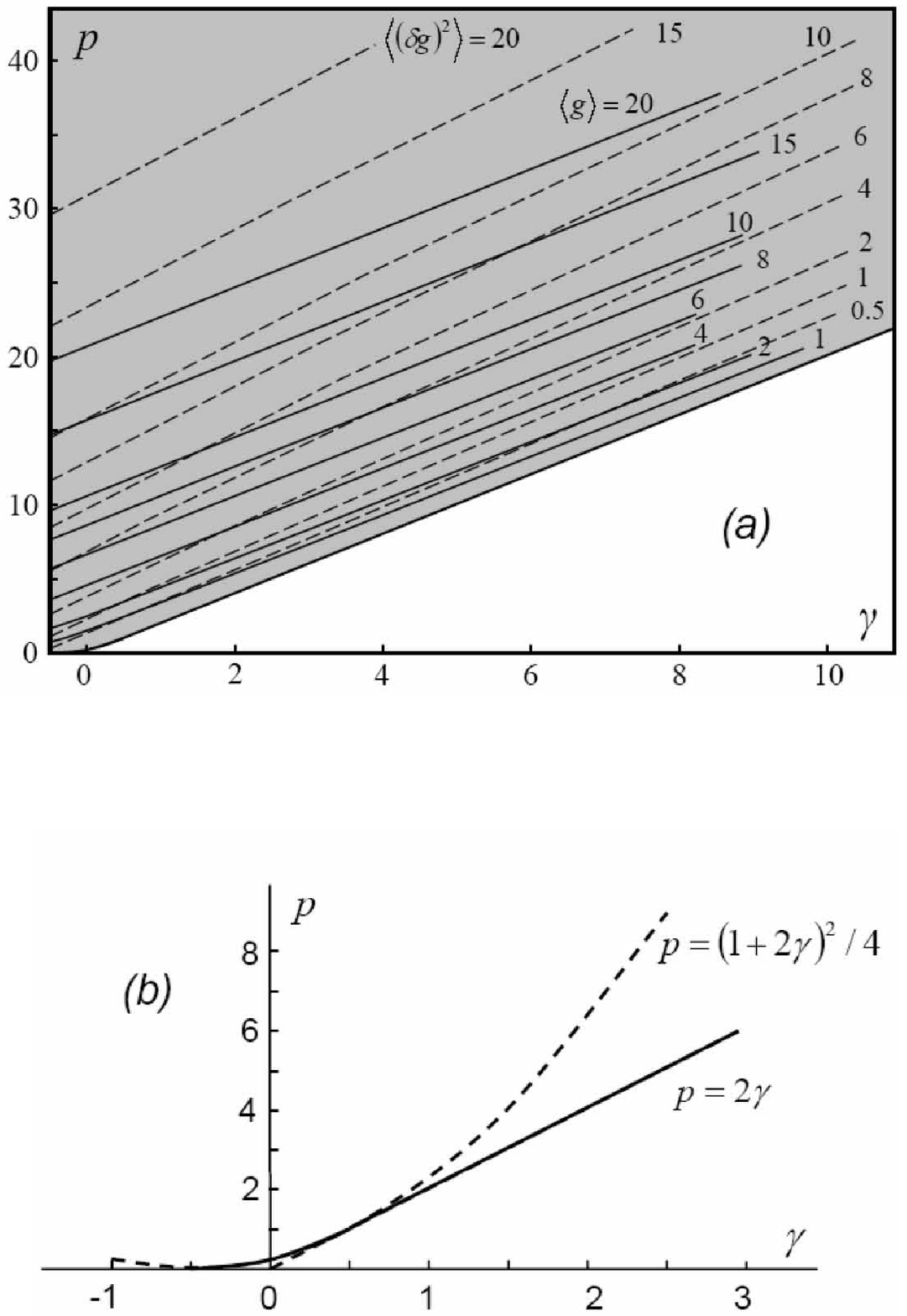}}
\caption{
(a) The phase diagram for $\tau_0=1$. Solid and dashed lines
correspond to constant values of $\langle g\rangle$
and  $\langle (\delta g)^2\rangle$.
(b) Construction, clarifying the arrangement of the lower
boundary.  }
\label{fig9}
\end{figure*}

The lower boundary of the physical region is determined by
two dependencies $p=2\gamma$ and $p=(1+2\gamma)^2/4$,
%which are matched
conjugated at the point $\gamma=0.5$ (Fig.9,b).
In approaching the lower boundary, coefficients $B_1$ and
$B_2$ tend to zero, maintaining condition $B_2\sim B_1$ for
$\gamma<0.5$, $B_2\sim B_1^{2\gamma}$ for $0.5<\gamma<1$
and $B_2\sim B_1^2$ for $\gamma>1$.

Indeed, due to positiveness of $B_2$ and negativeness of
$B_1$ the first relation (50) leads to the inequality
$$
p-2\gamma= |B_1|+2 B_2/|B_1|\ge 0\,,
$$
which gives a necessary condition $p\ge 2\gamma$ for
the physical solution. Setting  $p=2\gamma+\epsilon$,
we can rewrite (48) in the form
$$
\hat L u \equiv  \tau(\tau\!+\!1)  u''
-2\gamma\tau u' +2\gamma u =-\tau(\tau\!+\!1) u'^2-\epsilon u    \,.
\eqno(67)
$$
Equation $\hat L u=0$ has a solution $u_0(\tau)=A\tau$,
which allows to take the right-hand-side into account
%in the iterative manner
iteratively
$$
\hat L \delta u =-A^2 (\tau^2+\tau)-\epsilon A \tau   \,.
\eqno(68)
$$
For solvability of equation (68) its right-hand-side should be
orthogonal to solution $v_0(\tau)$ of the conjugated
equation $\hat L^+ v_0=0$
$$
0=(v_0,\hat L \delta u) =-A^2 (v_0,\tau^2+\tau)
-\epsilon A (v_0,\tau)\,,
\eqno(69)
$$
where $v_0(\tau)=(1+\tau)^{-1-2\gamma}$. For  $\gamma>1$,
the integrals corresponding to scalar products
$(v_0,\tau^2)$ and $(v_0,\tau)$ converge, so
$A\sim \epsilon$, $\delta u\sim \epsilon^2$. Since the linear
and quadratic in $\tau$ terms contain in  $u_0(\tau)$ and
$\delta u(\tau)$ correspondingly, then  $B_1\sim \epsilon$,
$B_2\sim \epsilon^2$, so $B_1$ and $B_2$ tend to zero in
approaching the line $p=2\gamma$, maintaining relation
$B_1^2\sim B_2$. For $\gamma<1$, the integral $(v_0,\tau^2)$
diverges and should be cut off at $\tau\sim 1/A$,
where transition to the logarithmic behavior (63) occurs;
as a result $A\sim \epsilon^{1/(2\gamma-1)}$, $\delta u\sim
\epsilon A\sim \epsilon^{2\gamma/(2\gamma-1)}$ and vanishing
of the coefficients occurs under condition $B_2\sim
B_1^{2\gamma}$.

The line  $p=(1+2\gamma)^2/4$ is distinguished due to the
fact, that a linearized version of (48) reduces to the
hypergeometric equation
$$
\tau(\tau+1) u''+(\alpha+\beta+1)\tau u' +\alpha
\beta u=0 \,,
\eqno(70)
$$
where parameters $\alpha$ and $\beta$ are given by
the formula
$$
\alpha, \beta = \frac{-(1+2\gamma)\pm \sqrt{(1+2\gamma)^2-4p}}{2}
\eqno(71)
$$
and become coinciding at the indicated curve. A solution
of (70), regular at the origin, has the following asymptotic
regimes
$$
u(\tau)=\left \{ \begin{array}{cc}
C\left[-{\scriptstyle\,\alpha\beta} \tau+
\frac{\alpha(\alpha+1)\beta(\beta+1)}{2}\,\tau^2+\ldots
 \right],& \tau\!\ll\! 1 \\
{    }\\
C\left[\frac{\Gamma(\alpha-\beta)}{\Gamma(\alpha)\Gamma(-\beta)}
\tau^{-\beta} +
\frac{\Gamma(\beta-\alpha)}{\Gamma(-\alpha)\Gamma(\beta)}
\tau^{-\alpha} \right], & \tau\!\gg \!1 \end{array}
\right. \,.
\eqno(72)
$$
Coefficients of  $\tau^{-\alpha}$ and $\tau^{-\beta}$
are determined by the matching conditions with (63),
so the combinations  $C\Gamma(\alpha-\beta)$ and
$C\Gamma(\alpha-\beta)$ remain finite in the limit
$\alpha\to\beta$. As a result, $C$ tends to zero due to
divergency of the gamma functions, and the coefficients of $\tau$
and $\tau^2$ (i.e. $B_1$ and $B_2$) disappear according
to the same law in approaching the curve $p=(1+2\gamma)^2/4$.
The given arguments are correct under condition that matching
with (63) occurs for sufficiently large $\tau$; practically
this condition is fulfilled for $\gamma<0.5$.

\begin{center}
{\small\bf 6.6. Metallic and dielectric regimes} \end{center}

According to \cite{3,4}, in the metallic state coefficients
$B_n$ obey  the same hierarchy $B_1\gg B_2 \gg B_3 \gg \ldots$
as for $d=2+\epsilon$; the estimate $B_n\sim p/\tau_0^n$
follows from Eqs.50 for large $p$ and
arbitrary $\gamma$, excluding a vicinity of the line $2\gamma=p$.
Substituting  the arising representation
$u(\tau)=p f(\tau/\tau_0)$ in the inverse Laplace transform,
one can  expand $f(x)$ in a series and retain two first
terms; it leads to the Gaussian distribution. The case
$2\gamma\approx p$ can be considered separately and leads to
the same conclusion (see Appendix 2).

The region of small $p$ can be analyzed rigorously for
$\gamma=-1/2$.  Linearizing (48) and omitting the small term $p
u$, we have a solution
$$
u(\tau)= A \ln\left(1+\tau/\tau_0\right) \,,
\eqno(73)
$$
tending to zero for  $\tau\to 0$. If $A$ is sufficiently small,
this solution remains valid in the region $\tau\gg \tau_0$,
where $\tau(\tau+\tau_0)$ can be replaced by $\tau^2$. After it,
equation (48) with  $\gamma=-1/2$ has the exact solution
$$
u(x)=-{\textstyle
 \frac{1}{4}} p(\ln{\tau}-x_0)^2 +{\textstyle \frac{1}{2}} \,.
\eqno(74)
$$
Within the accepted accuracy, one can replace $\tau$
by $\tau\!+\!\tau_0$ and provide the condition $u(0)=0$ by
the appropriate choice of $x_0$. Then the solution
$$
u(x)=- (p/4)\ln^2(\tau+\tau_0)  -\mu\, \ln(\tau+\tau_0)
+b
\eqno(75)
$$
with arbitrary $\mu$ and $b=1/2-\mu^2/p$,
$\tau_0=\exp\left(\sqrt{2/p}-2\mu/p\right)$ is valid for all
$\tau$; it is in agreement with (73) and for $\mu\ll 1$ provides
a suggested smallness of $A$.  Eq.75 corresponds to the
log-normal distribution; it is evident for the saddle point
calculation, but needs a more subtle analysis for small $p$,
where applicability of the saddle-point method is strongly
restricted (see Appendix 2).  As demonstrated in the same
Appendix, the approximately log-normal distribution remains valid
for arbitrary values of $\gamma$ in the region of small $p$.

It is clear from above considerations, that large $p$
correspond to the metallic state, and small $p$ refer
to the localized
regime.  Since parameters $p$ and $\gamma$ are in the one-to-one
correspondence with $L/\xi$ and $L_z/L$, the certain line
in the $(p\,,\gamma)$ plane corresponds to the cubical
systems. This line originates at the point $p=0$, $\gamma=-1/2$
and goes to the region of large  $p$.

\begin{figure*}
\centerline{\includegraphics[width=5.1 in]{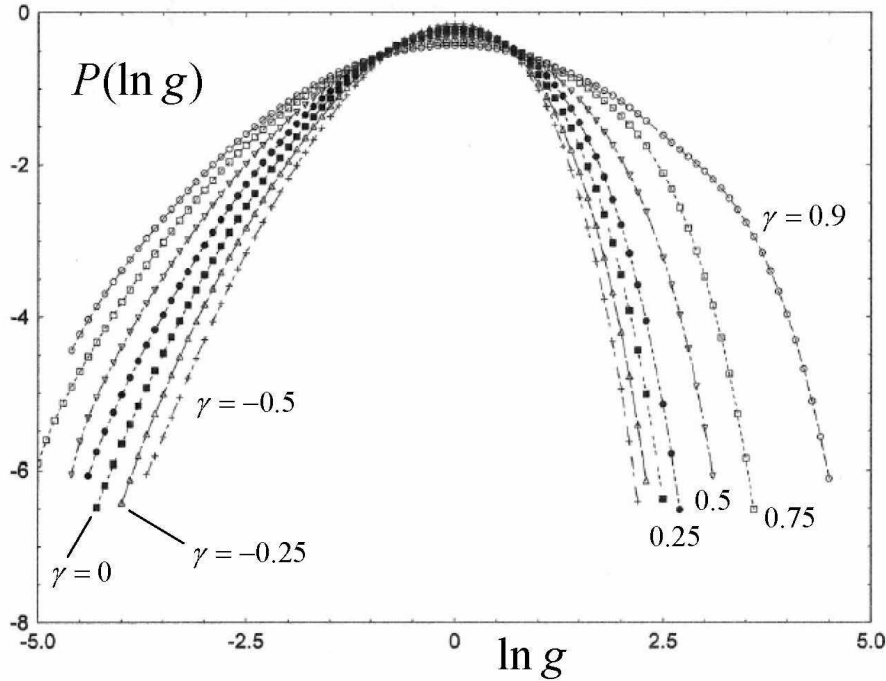}}
\caption{
Distribution $P(\ln g)$ for different $\gamma$ and a fixed
value $p=2$.  }
\label{fig10}
\end{figure*}

\begin{center}
{\small\bf 6.7. Critical region}
\end{center}

The critical region corresponds to values $p\sim 1$, when
it is necessary to  solve Eq.48 numerically, and
 numerically realize the inverse Laplace
transformation.  Integration of (48) begins at large $\tau$ in
the logarithmic coordinates, starting from the asymptotics (63),
and continues for $\tau\alt 1$ in the usual coordinates, adjusting
$x_0$ to the condition $u(0)=0$. It is essential to control
monotonicity of the solution,  not allowing oscillations, which
are possible according to Fig.7,b.

A numerical realization of the inverse Laplace
transformation is conveniently made, using the rational
approximation of $F(\tau)$ with subsequent
decomposition to simple fractions
$$
F(\tau)=\frac{P_M(\tau)}{Q_N(\tau)} \,=\,
\sum\limits_{i=1}^{N} \,\frac{A_i}{\tau-\tau_i}  \,,
\eqno(76)
$$
where  $P_M(\tau)$ and $Q_N(\tau)$ are polynomials of degree
$M$ and  $N$, and $M$ should be smaller than $N$ to avoid
$\delta$-functional contributions at the origin;
after it, $W(g)$ is represented in the form
$$
W(g)=\sum\limits_{i=1}^{N}\, A_i \,\exp\{\tau_i g\} \,.
\eqno(77)
$$
Due to decreasing of $W(g)$ for $g\to\infty$,
the function  $F(\tau)$
cannot contain singularities in the right half of the
complex plane.  In practice, the poles with a positive real
part may arise due to
%the so called
 "defects", manifested as
%in appearance of
pairs of the close pole and root: it leads
to a catastrophic loss of accuracy or overflow. A practical
recipe consists in the use of approximants of the maximal order,
not containing the poles with a positive real part and providing
the maximal accuracy of approximation for $F(\tau)$. A situation
with "defects" can be optimized by changing the number and
location of points on the $\tau$ axis, which are used for
approximation (76).

Fig.10 demonstrates distributions $P(\ln g)$ for different
$\gamma$ and a fixed value $p=2$.
In a vicinity of the line $2\gamma=p$,
a distribution is practically symmetric and close to the
log-normal one. With decreasing of $\gamma$, asymmetry of the
distribution arises; at first it increases quickly, and then
remains practically on the same level. Comparison with numerical
data by Markos  \cite{10} is represented in Fig.11; agreement is
satisfactory for parameters $\gamma=0$, $p=0.85$.  Variation of
the parameter $\tau_0$ does not change  the form of the
distribution $P(\ln g)$ and leads only to its parallel shift. For
agreement with \cite{10} one should set $\tau_0=0.67$, i.e. the
 choice $\tau_0\sim 1$ provides a correct scale of conductance in
the critical region; it corresponds to expectations from the
results for $d=2+\epsilon$.

As was indicated in the end of Sec.6.6, the trajectory
corresponding to cubical systems  exists in the $(p,\gamma)$
plane; along it the form of the distribution $P(\ln g)$
is modified due to a change of the amplitude of
a random potential.
A state of the system can be characterized by two
parameters, a position on the trajectory and a value of $\tau_0$.
In order the whole distribution obeys  one-parameter scaling,
it is sufficient and necessary that such scaling (i.e. dependence
only on $L/\xi$) was valid for two independent parameters,
characterizing this distribution, e.g. for $\langle g\rangle$
and $\langle (\delta g)^2\rangle$. Such property for the first
parameter was established in \cite{23} in the framework of
self-consistent theory of localization \cite{100}; its
validity for the second parameter looks rather probable
due to results for the metallic  ($\left\langle
(\delta g)^2 \right\rangle=const$) and localized
($\left\langle (\delta g)^2 \right\rangle\sim \exp(const L/\xi)$)
phases.

According to Fig.11, a difference of the theoretical
curve from the data by Markos \cite{10} reduces to
smearing of a singularity at point $A$ (Fig.2). This difference
is not surprising. In the present paper we use the invariant
definition of conductance, independent of the way
how the contact resistance of the reservoir is excluded, and
certainly characterizing a finite system (Sec.4); in such
situations singularities are impossible in accordance with the
general principles \cite{12,13}. The definition used
in \cite{10} is given by the many-channel
Economou--Soukoulis formula \cite{26,27}, which contains built-in
singularities: the distribution $W(g)$ for each channel drops off
abruptly at $g=1$.  This defect is related with the unsolved
problem of the contact resistance and would be inessential in the
true many-channel situation, when each channel carries a small
part of conductance.  In fact, the analysis by Markos shows
\cite{10}, that the critical distribution is determined mainly by
the most transparent channel, while the rest of channels forms
only the exponential tail; so a deficiency of the definition is
essential and directly leads to a singularity.  On the other
hand, absence in \cite{10} of semi-transparent boundaries,
separating the system from ideal leads, results in its strong
interaction with the environment. The thermodynamic limit,
realized by increasing the size of ideal leads, refers to the
composite system "sample+ideal leads" and allows existence of
singularities. In conclusion, a singularity at point $A$ is quite
possible in the framework of the calculational scheme of
\cite{10}, but is surely related with a deficient definition of
conductance.

\begin{figure}
\centerline{\includegraphics[width=2.9 in]{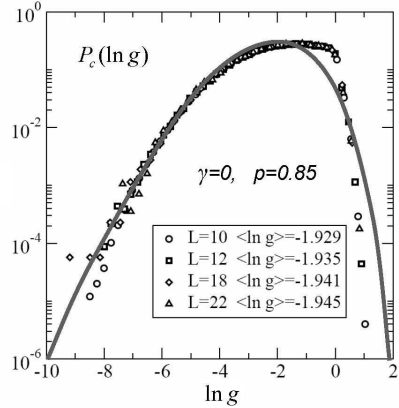}}
\caption{
Comparison of numerical data by Markos \cite{10} with
the results of present theory for $\gamma=0$, $p=0.85$.  }
\label{fig11}
\end{figure}

\begin{center}
{\bf 7. Conclusion}
\end{center}

Above we have shown, that a modification of the
Shapiro approach allows to
introduce the two-parameter family of conductance
distributions, defined by simple differential equations,
whose parameters $p$ and $\gamma$ are in the
one-to-one correspondence with parameters $L/\xi$ and $L_z/L$,
characterizing a quasi-one-dimensional system. We did not
try to calculate these parameters for any specific situations
but investigate all family of distributions in whole.
For large $p$ all distributions are Gaussian, which is typical
for the metallic state. For small $p$ distributions are close to
log-normal, in accordance with expectations for the
localized phase. For $p\sim 1$ and the $\gamma$ values  in the
left part of the phase diagram, distributions are highly
asymmetrical and close to the one-sided log-normal distribution,
as established in numerical experiments for the critical region.
 For a certain choice of parameters, we reproduce the results for
 cumulants $\left\langle\!\left\langle g^n
\right\rangle\!\right\rangle$ in the space dimension
$d=2+\epsilon$ obtained in the framework of the
nonlinear $\sigma$-model approach.
Numerical results for the critical distribution at $d=3$
are reproduced satisfactorily for $\gamma=0$, $p=0.85$ (Fig.11),
excluding a singularity at point $A$ (Fig.2).  The
latter  singularity is admissible in the framework of
calculational scheme, used in \cite{10},  but  related with
a deficient definition of conductance.

 The universal property of distributions is existence
of two asymptotic regimes, log-normal for small $g$ and
exponential for large $g$, while their actuality depends on
a specific situation.  In the metallic phase, a distribution
is determined mainly by the central Gaussian peak, while
two indicated asymptotic regimes refer to remote tails. In the
critical region these two regimes determine practically all
distribution, and the log-normal behavior extends to the region of
the maximum. In the localized phase,  the latter behavior extends
even more and forces out the exponential  asymptotics to the
region of a remote tail.

The assumption of one-parameter scaling is the basis
%base
%lies in the foundation
of the analysis, and its results are in agreement with this
assumption. One-parameter scaling for the whole distribution
takes place under condition, that two independent parameters
characterizing this distribution, are functions of the ratio
$L/\xi$. Such property is established in  \cite{23} for $\langle
g\rangle$  and looks rather probable for
$\langle (\delta g)^2\rangle$.

The present paper fills in one of the main gaps in theory of
disordered systems, related with absence of the systematic
methods for investigation of distributions.
%At present, our
%conception leading to Eq.47 looks  as completely justified,
%and the problem of the conductance distribution can be
Our conception, leading to Eq.47, is a natural consequence of
one-parameter scaling and looks as completely justified;
so the problem of the conductance distribution can be
considered as solved in principle. The remaining free parameters
can be fixed by calculation of the first several moments of
conductance, which can be made by the standard methods.

\begin{center}
{\it Appendix 1.} Derivation of the evolution equation
\end{center}

The increase of the length of a 1D system from $L$ till
$L+\Delta L$ is assisted by multiplication of transfer-matrices,
$\hat T_{L+\Delta L}=\hat T_{L} \hat T_{\Delta L}$. Let assume
a form (16) for the matrix $\hat T_{L}$ and use the following
representation for the matrix $\hat T_{\Delta L}$
$$
\hat T_{\Delta L}=
 \left ( \begin{array}{cc} \sqrt{1\!+\!\epsilon^2}\, e^{i\beta_1}
 & -i\epsilon \,e^{i\beta_2} \\ i\epsilon\, e^{-i\beta_2} &
\sqrt{1\!+\!\epsilon^2}\, e^{-i\beta_1} \end{array} \right)\,,
\eqno(A.1)
$$
where $\epsilon$, $\beta_1$, $\beta_2$ are small random
quantities. The analogy with a point scatterer shows
(see (26)), that $\epsilon$ is proportional to the
amplitude of the random potential and its average
should be set to zero, since in the other
case it can be achieved by a change of the energy
origin. Multiplying matrices, one obtains for the parameter
$\tilde\rho$, corresponding to the matrix $\hat T_{L+\Delta L}$,
in the second order in $\epsilon$
$$
\tilde \rho=\rho-2\epsilon\sqrt{\rho(\rho+1)} \sin\psi
+\epsilon^2 (2\rho+1)\equiv f(\rho)\,,
\eqno(A.2)
$$
where
$$
\psi=\theta-\varphi+\beta_1+\beta_2\,.
\eqno(A.3)
$$
For the distribution of $\tilde\rho$ we have
$$
P_{L+\Delta L}(\tilde\rho) = \int\,d\rho\, d\psi\,d\epsilon
P_{L}(\rho) P(\psi) P(\epsilon)
\delta\left(\tilde\rho-f(\rho)\right)=
$$
$$
=\int\, d\psi\,d\epsilon  P(\psi) P(\epsilon)
P_{L}\left(f_1(\tilde\rho)\right) f'_1(\tilde\rho) \,,
\eqno(A.4)
$$
where $\rho=f_1(\tilde\rho)$ is the inverse function to
$\tilde\rho=f(\rho)$, which is found by iterations in
$\epsilon$
$$
f_1(\rho)=\rho+2\epsilon\sqrt{\rho(\rho+1)} \sin\psi
+
$$
$$
+\epsilon^2(2\rho+1)\left(2 \sin^2\psi-1\right)
\,.
\eqno(A.5)
$$
Substituting to $(A.4)$ and expanding to the second order
in $\epsilon$, we have
$$
P_{L+\Delta L}(\rho) = P_{L}(\rho)
\left[1+2\,\overline{\epsilon^2}
\left(2\,\overline{\sin^2\psi}-1\right)\right] +
$$
$$
+ P'_{L}(\rho)\, \overline{\epsilon^2} (2\rho\!+\!1)
\left(4\,\overline{\sin^2\psi}-1\right)
+
$$
$$
+P''_{L}(\rho) \,\overline{\epsilon^2} \,\rho(\rho\!+\!1)
2\,\overline{\sin^2\psi}  \,,
\eqno(A.6)
$$
and setting  $\overline{\epsilon^2}=\alpha\Delta L$,
$\overline{\sin^2\psi}=A_0$, we come to (14) with parameters
(18), if small quantities $\beta_1$ and  $\beta_2$ are neglected
in $(A.3)$.

\begin{center}
{\it Appendix 2.} To investigation of Eq.48
\end{center}

Let fill in the gaps in investigation of Eq.48
%some technical details omitted in the main text.
allowed in the main text.

\vspace{3mm}

{\it A vicinity of the line $p=2\gamma$.} In the case
$p=2\gamma$, the linearized in $u$ equation (48)
has an exact solution $u_0(\tau)=A\tau$  with small $A$,
which is extended to the region of large $\tau$, where
Eq.48  reduces to (60) after the change $x=\ln\tau$.
Considering $p\gg 1$ and retaining the main in $p$ terms,
we have the equation
$$
 u_{x}'^2-p u_x' +p u =0 \,,
\eqno(A.7)
$$
whose solution can be written in the parametric form \cite{28}
$$
x=\ln\tau=t+\ln t+ x_0\,,\qquad u=-{\textstyle \frac{1}{4}} p
(t^2+2t) \,,
\eqno(A.8)
$$
where the running parameter $t$ changes from zero to
infinity. Producing the inverse Laplace transformation
and changing from integration over $\tau$ to integration
over $t$, we have
$$
W(g) = \frac{1}{2\pi i} \int \,dt (t+1) e^{t+x_0} \cdot
$$
$$
\cdot
\exp\left\{-\frac{1}{4} p\,(t^2+2t)  + gt e^{t+x_0}
\right\} \,, \eqno(A.9)
$$
and the use of the saddle-point approximation leads to the
log-normal distribution
$$
W(g) =  \sqrt{\frac{p}{4\pi}} \frac{1}{g}
 \exp\left\{- \frac{p}{4} \left(\ln{\frac{g_c}{g}} \right)^2
 \right\} \,,
$$
$$
 \qquad g_c=\frac{p}{2} e^{-x_0} \,,
 \eqno(A.10)
$$
reducing to the Gaussian one in the case of large  $p$.

\vspace{3mm}

{\it Small $p$.} For $\gamma=-1/2$, the solution $u(\tau)$
is determined by Eq.75, whose substitution to the
inverse Laplace transform and
%successive
subsequent
changes $\tau\to \tau\!-\!\tau_0$ and $\tau\to \tau/g$ give
$$
W(g) = \frac{1}{2\pi i} g^{-1+\mu}
\int\limits_{-i\infty}^{i\infty} \,d\tau \tau^{-\mu} e^\tau
\cdot
$$
$$
\cdot
 \exp\left\{-\frac{p}{4} \ln^2\tau +\frac{p}{2} \ln\tau \ln g
 -\frac{p}{4} \ln^2g +b -\tau_0 g  \right\} \,.
  \eqno(A.11)
$$
Expanding the exponent in $p\ln\tau$ and calculating
the integrals for small $\mu$, we have
$$
W(g) \sim \frac{1}{g}
 \exp\left\{- \frac{p}{4} \left(\ln{g}-\frac{2\mu}{p}
 +\frac{1}{\mu} \right)^2 -\tau_0 g
 \right\} \,.
 \eqno(A.12)
$$
and the term $\tau_0 g$ can be omitted for $\mu\ll \sqrt{p}$.
Consideration is valid for $p|\ln g|\ll 1$, which in
the case  $p\ll \mu\ll\sqrt{p}$ covers a vicinity of the maximum
and describes all essential part of the distribution.

In fact, the approximately log-normal distribution
is valid in the region of small $p$ for the arbitrary
value of $\gamma$.  Indeed, setting $\gamma=-1/2+\epsilon$
(where
$0\le\epsilon\le\sqrt{p}$) and omitting the term $pu$
in the linearized equation (48), we have the solution
$$
u(\tau)= C_1 (\tau+\tau_0)^{2\epsilon} +C_2 \,,
\eqno(A.13)
$$
which reduces to (73) for $\epsilon \ln(\tau+\tau_0) \ll 1$.
On the other hand, Eq.60 with
$\tilde\gamma=2\epsilon$ has an approximate solution
for large $x$
$$
 u(x)=- {\textstyle \frac{1}{4}} p x^2 + \epsilon x(\ln{x}-1)
 - (\epsilon^2/p)
 \ln^2{x}+{\textstyle \frac{1}{2}}
+ O(x^{-1})\,,
 \eqno(A.14)
$$
where $O(x^{-1})$ contains terms of type $\ln^m{x}/x$.
Neglecting slow variation of  $\ln x$ and replacing
it by a suitable constant, one can see that
invariance relative $x\to x-x_0$ allows to reduce $(A.14)$ to a
form (75).

%\newpage

\end{document}